\documentclass[fleqn,10pt]{wlscirep}
\usepackage[utf8]{inputenc}
\usepackage[T1]{fontenc}

\usepackage{amsmath}
\usepackage{multicol}
\usepackage{listings} 
\usepackage{hyperref}

\title{Evolutionary Random Graph for Bitcoin Overlay and Blockchain Mining Networks}

\author[1,*]{Jacques Bou Abdo}
\author[1]{Shuvalaxmi Dass}
\author[1]{Basheer Qolomany}
\author[1]{Liaquat Hossain}
\affil[1]{University of Nebraska at Kearney, Cyber Systems, Kearney, NE, USA}

\affil[*]{bouabdoj@unk.edu}


\begin{abstract}
The world economy is experiencing the novel adoption of distributed currencies that are free from the control of central banks. Distributed currencies suffer from extreme volatility, and this can lead to catastrophic implications during future economic crisis. Understanding the dynamics of this new type of currencies is vital for empowering supervisory bodies from current reactive and manual incident responders to more proactive and well-informed planners. Bitcoin, the first and dominant distributed cryptocurrency, is still notoriously vague, especially for a financial instrument with market value exceeding \$$1$ trillion. Modeling of bitcoin overlay network poses a number of important theoretical and methodological challenges. Current measuring approaches, for example, fail to identify the real network size of bitcoin miners. This drastically undermines the ability to predict forks, the suitable mining difficulty and most importantly the resilience of the network supporting bitcoin. In this work, we developed Evolutionary Random Graph, a theoretical model that describes the network of bitcoin miners. The correctness of this model has been validated using simulated and measure real bitcoin data. We then predicted forking, optimal mining difficulty and network size.
\end{abstract}
\begin{document}

\flushbottom
\maketitle
%
%
\thispagestyle{empty}

\section*{Introduction}\label{sec1}

The advent of scale-free networks \cite{barabasi_emergence_1999, bollobas_diameter_2004, albert_diameter_1999} evolved attempting to explore the viability of Erdös and Rényi’s random graph \cite{erdos_random_1959}  within the context of real-world experimental data. Real-world information networks such as bitcoin overlay networks, the network of miners supporting all bitcoin transactions, do not follow the scale-free characteristics. A number of attempts, using physical probing, to map bitcoin overlay networks failed due to constraints such as failing to reach the nodes behind proxies and distinguishing clusters of nodes \cite{lischke_analyzing_2016,deshpande_btcmap_2018, neudecker_network_2019, goldberg_txprobe_2019,essaid_visualising_2019,ben_mariem_vivisecting_2018, eisenbarth_open_2021, bohme_bitcoin_2014, park_nodes_2019, Miller2015DiscoveringB}, which makes it an open research question since $2014$. To overcome this, we propose evolutionary random graph, a theoretical model describing the bitcoin overlay network, that helps in predicting the bitcoin network topology, correcting existing mapping and predicting its emerging behavior. Our results suggest that the bitcoin overlay network's equilibrium size is proportional to bitcoin's price.

\subsection{Bitcoin overlay network}\label{sec1.1}
The bitcoin network can be represented as a stack of four layers, as shown in figures \ref{bitcoin_stack_transaction}, \ref{bitcoin_stack_bitovernet}, \ref{bitcoin_stack_p2p} and \ref{bitcoin_stack_internet}. The lowest layer is the internet itself (although bitcoin can be hosted in a private network), which is responsible for the physical exchange of messages. The second layer is the overlay network which is a virtual topology created on top of the physical network to abstract the underlying network in such a way that physical distances become meaningless. In figure \ref{bitcoin_stack_internet}, the underlying network resembles part of the internet, while the overlay network, figure \ref{bitcoin_stack_p2p}, creates links (neighboring links) and neighborhoods, represented as red dotted lines, irrespective of the underlying network structure. Since the links are not restricted by a maximum distance, overlay networks are free from the spatial-temporal restrictions imposed on the internet. The third layer is the Bitcoin Overlay Network (BitOverNet) which divides the nodes into two types. The first type is the client/wallet (represented as green node in figure \ref{bitcoin_stack_bitovernet}), who requests a transaction through a miner. The second type is the miner (represented as red node in figure \ref{bitcoin_stack_bitovernet}) who receives the request from his/her clients and inform other miners about the received request before competing on generating the new block of transactions. Since each client is connected to one miner and not connected to other clients, the BitOverNet can be defined as the network of miners (represented by red nodes and yellow links in figure \ref{bitcoin_stack_bitovernet}). The network of miners, the focus of our study, is the backbone for all bitcoin transactions. The highest layer is the bitcoin transaction network, figure \ref{bitcoin_stack_transaction}, which is a virtual network representing the clients involved in bitcoin transactions. The network is virtual since the clients are not connected using any communication channel, but the clients’ ids are mentioned in the same transaction (like the foreign key relationships in relational databases). The miners are transparent at this layer and thus considered a network of clients and transactions.

\subsection{Mapping the bitcoin overlay network}\label{sec1.2}
In bitcoin networks, miners follow a competing/competitive process in generating a valid bitcoin block. When a block is generated, it gets broadcasted to the successful miner’s neighbors who verify and rebroadcast it until all the miners gets informed, similar to information diffusion \cite{noauthor_fake_2017, rogers_innovation_nodate}, contagious diseases \cite{cirillo_tail_2020} and other social contagious problems \cite{giles_social_2011}. This operation is called consensus and requires convergence time. If two miners independently issue two blocks within an interval shorter than convergence time, the network fails to reach consensus and the bitcoin forks into two chains, as shown in figure \ref{forking_fig}. More on bitcoin forking and its rate can be seen in \cite{neudecker_short_2019}. The system can recover later by selecting the longer blockchain (this is one of the reasons participating in the probabilistic finality of the bitcoin transaction). Although the system can recover later, but this has dramatic consequences on the bitcoin users, such as double spending \cite{karame_misbehavior_2015}.

The rate of block issuance is function of the miners’ computation power (proportional) and mining difficulty (inversely-proportional) i.e. $\lambda=f($\textit{computational-power, mining-difficulty}$)$. The convergence time is function of the network diameter i.e. $T_{conv}=f($\textit{network-diameter}$)$. Forking probability thus follows Poisson distribution with rate $\lambda$ and period $T_{conv}$. To decrease the forking probability, we can either:
\begin{enumerate}
  \item \textbf{Increase the mining difficulty}: This is possible, but risks making mining financially infeasible. Currently, this is the only used option.
  \item \textbf{Decrease the computational power}: The computational power is pseudo-following Moore’s law and thus always increasing (and that’s why mining difficulty gets revisited regularly to counter the increase in computational power). So, this option is not feasible.
  \item \textbf{Decrease the network diameter}: this would either require decreasing the number of nodes or changing the network structure (changing its scalability). In both cases, we need to first measure the existing network before recommending any changes. The implications of changing the network diameter are shown in figure \ref{sys_dynamics_fig}.
\end{enumerate}

\begin{figure}[ht]
\centering
\includegraphics[width=0.75\textwidth]{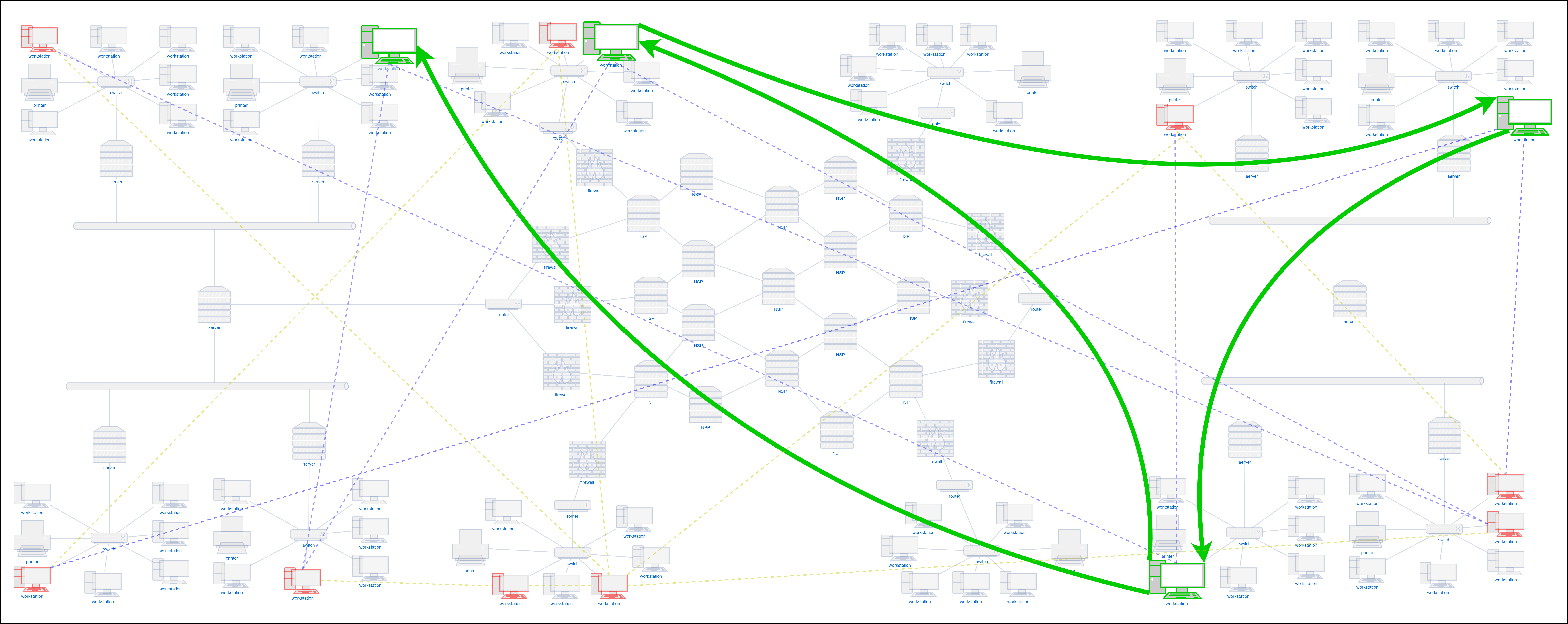}
\caption{Bitcoin transaction network}
\label{bitcoin_stack_transaction}
\end{figure}

\begin{figure}[ht]
\centering
\includegraphics[width=0.75\textwidth]{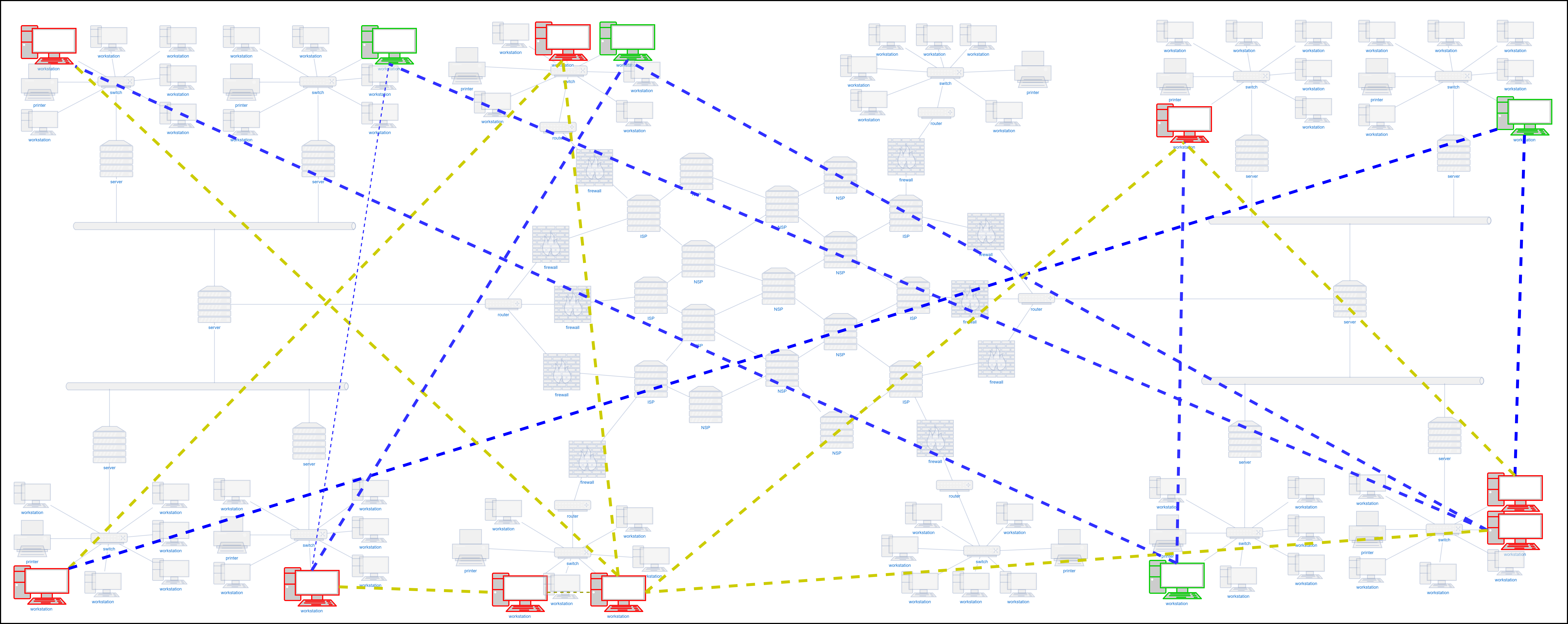}
\caption{Bitcoin overlay network (the focus of this study)}
\label{bitcoin_stack_bitovernet}
\end{figure}

\begin{figure}[ht]
\centering
\includegraphics[width=0.75\textwidth]{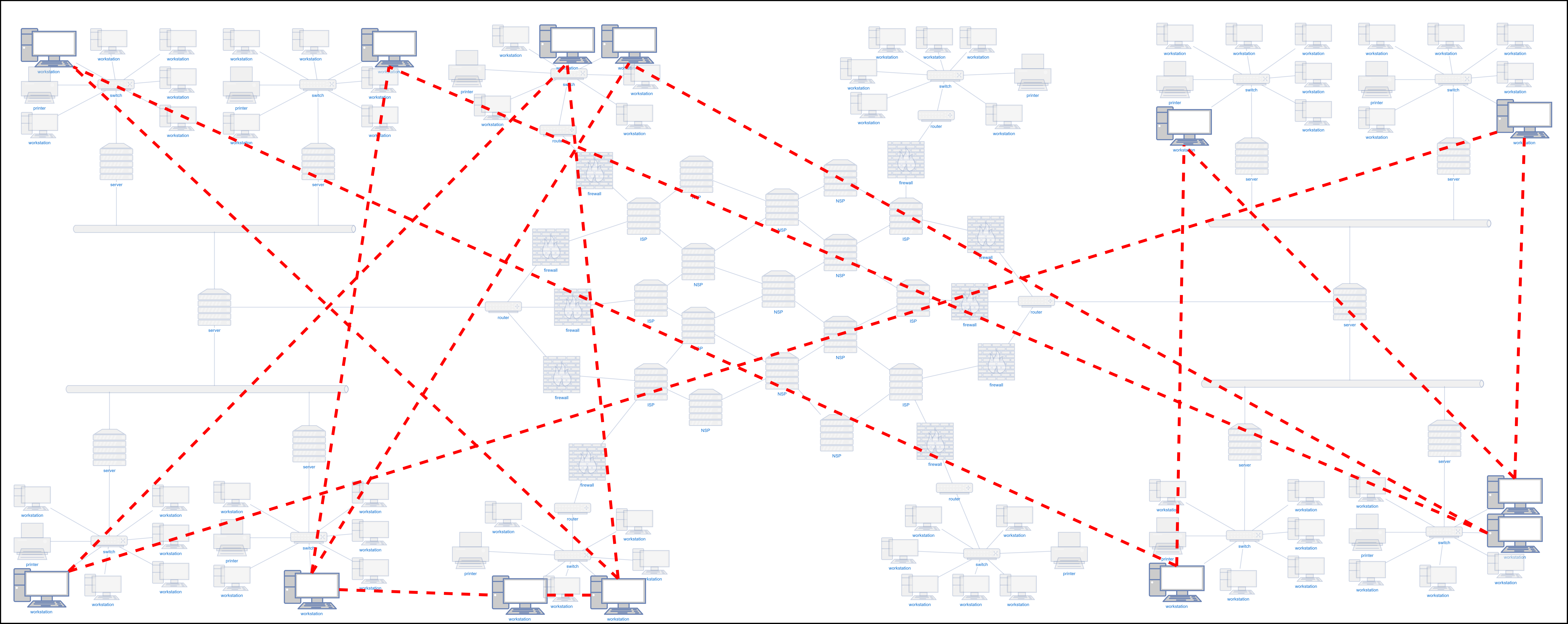}
\caption{Overlay network (peer-to-peer)}
\label{bitcoin_stack_p2p}
\end{figure}

\begin{figure}[ht]
\centering
\includegraphics[width=0.75\textwidth]{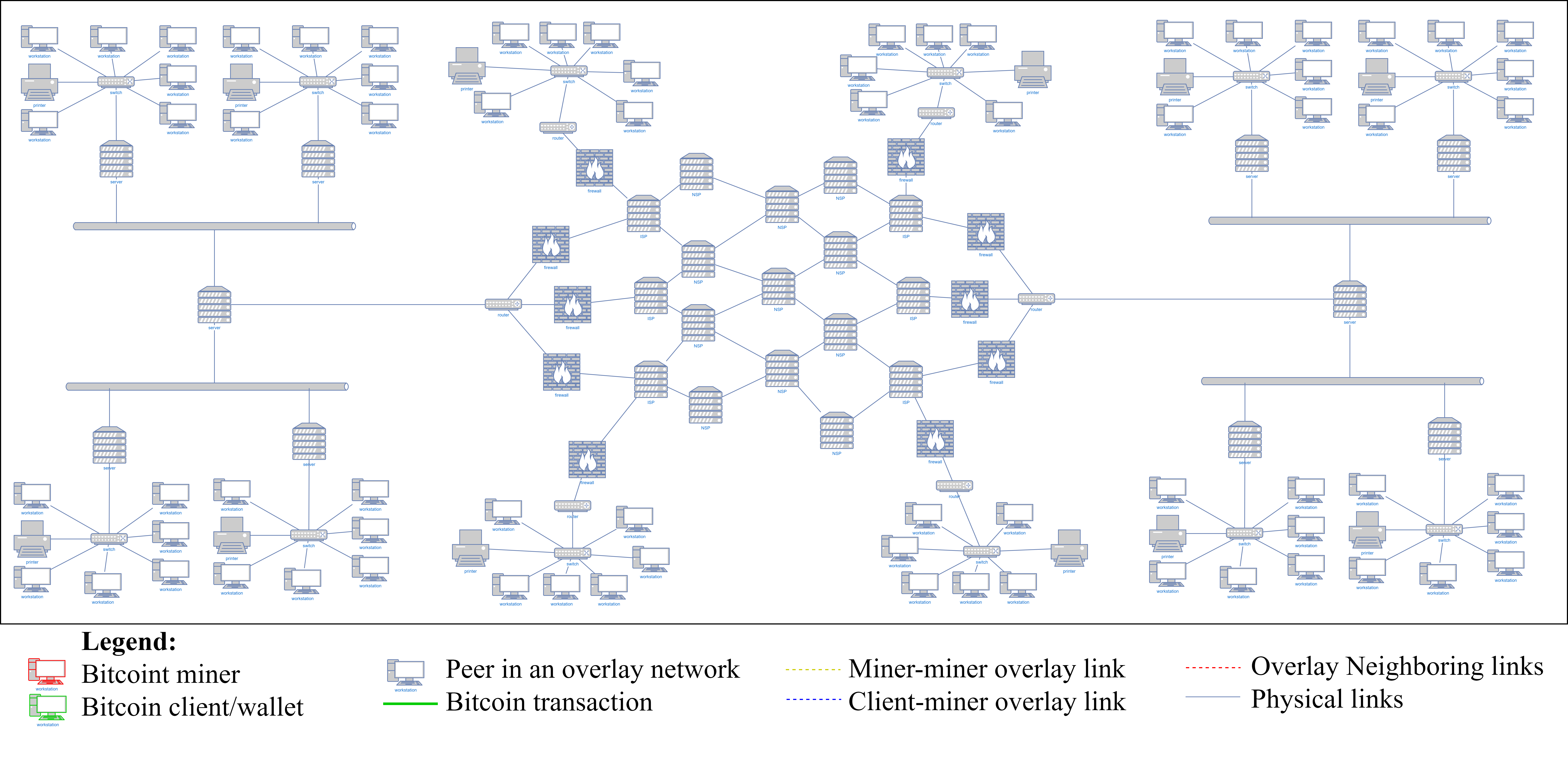}
\caption{Physical network (Internet)}
\label{bitcoin_stack_internet}
\end{figure}

\begin{figure}[ht]
\centering
\includegraphics[width=0.75\textwidth]{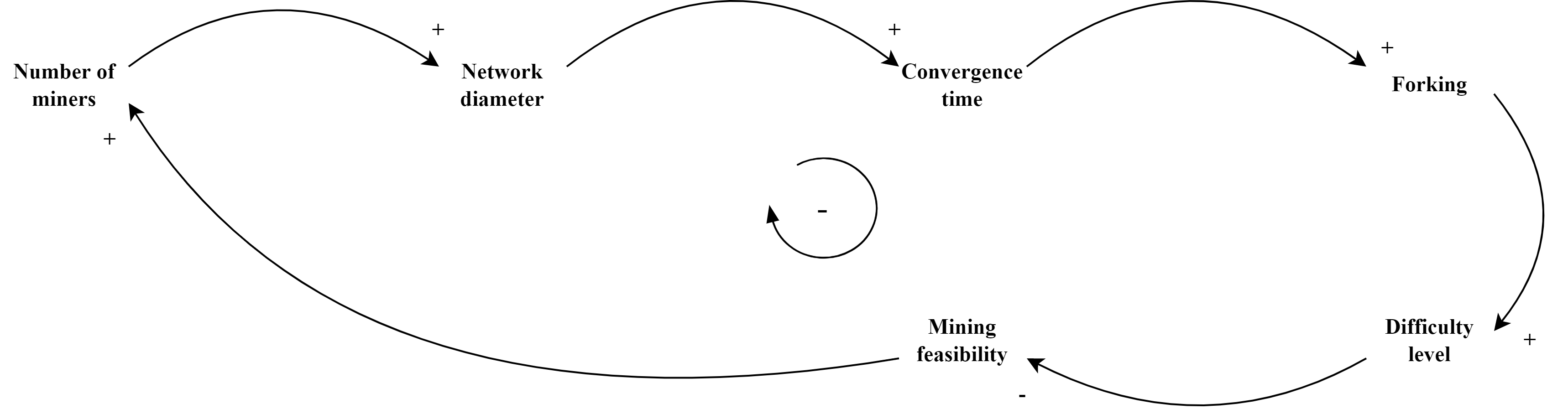}
\caption{Bitcoin network equilibrium: Implications of network diameter}
\label{sys_dynamics_fig}
\end{figure}

\begin{figure}[ht]
\centering
\includegraphics[width=0.75\textwidth]{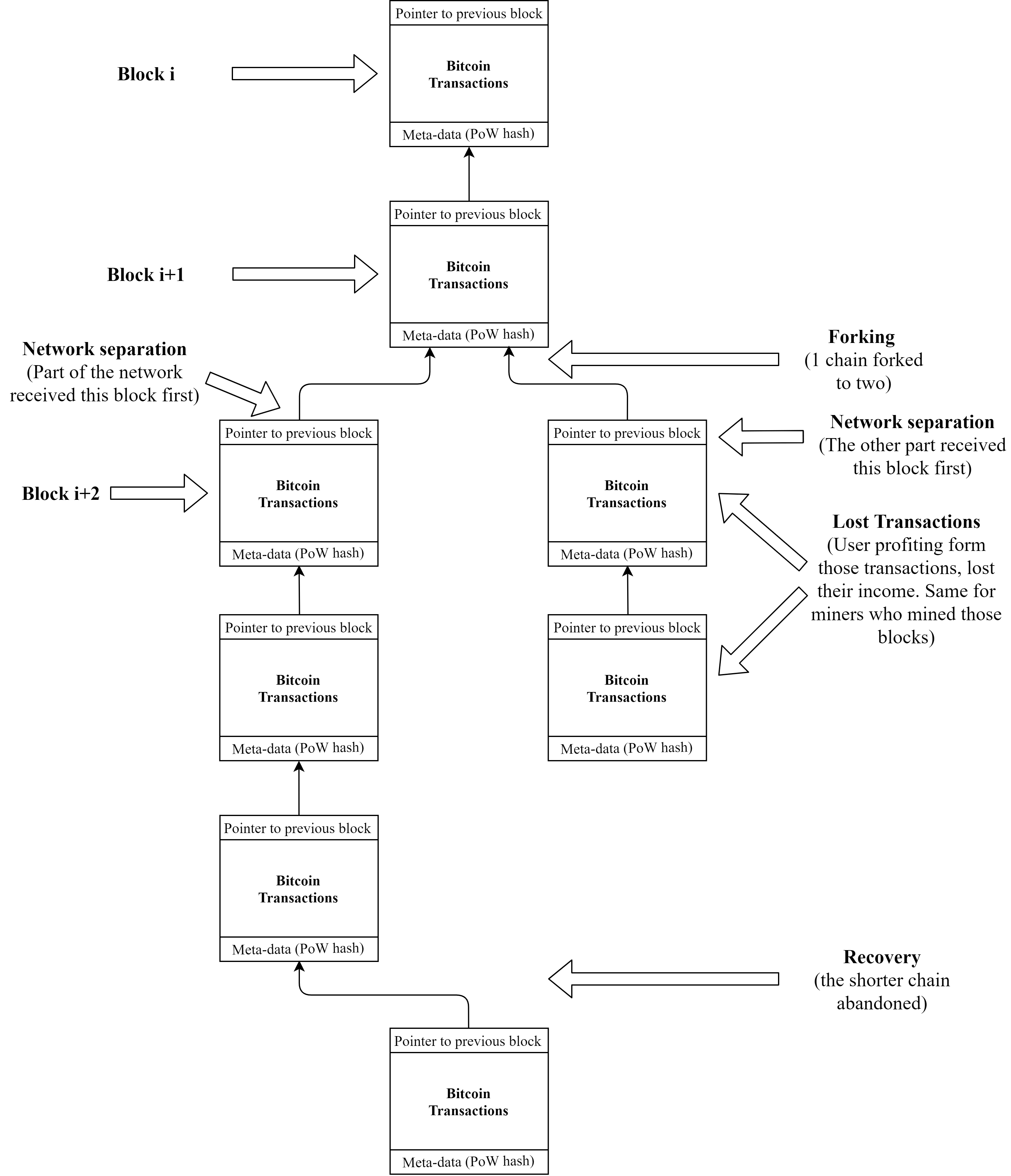}
\caption{Bitcoin's blockchain, forking and recovery}
\label{forking_fig}
\end{figure}

Knowing the network diameter is vital in understanding the BitOverNet, predicting forks and intervening to prevent crashes from happening. Mapping it also has implications on improving its security such as its resilience against network attacks \cite{saad_partitioning_2019, tran_routing-aware_2021, tran_stealthier_2020} and robustness against selfish \cite{bai_deep_2018}, trail \cite{grunspan_profitability_nodate} and stubborn miners \cite{nayak_stubborn_2015}. As discussed previously, scientists failed in mapping the BitOverNet \cite{lischke_analyzing_2016,deshpande_btcmap_2018, neudecker_network_2019, goldberg_txprobe_2019,essaid_visualising_2019,ben_mariem_vivisecting_2018, eisenbarth_open_2021, bohme_bitcoin_2014, park_nodes_2019, Miller2015DiscoveringB} and this hinders progress in protecting it. It is worthy of noting that mapping the BitOverNet, figure \ref{bitcoin_stack_bitovernet}, is radically different than the bitcoin transaction network \cite{serena_cryptocurrencies_2021, tao_complex_2021, fischer_complex_2021, michalski_dru_2020}, figure \ref{bitcoin_stack_transaction}, and bitcoin lightning \cite{bartolucci_percolation_2020}.

\section{Evolutionary Random Graph}\label{sec2}
One of the first theorized networks is the random graph by Erdös and Rényi \cite{erdos_random_1959} where it is generated by having $N$ available nodes equiprobably connected. For example, each two nodes are connected, using a link, with a probability $p$, as shown in algorithm \hyperlink{Algorithm_1}{1} at table \ref{tabtable_1}. When node $(N+1)$ gets introduced, it gets connected to each other node with the same probability $p$. The number of nodes having $k$ links nicely follows Poisson distribution, but as shown by Albert and Barabasi \cite{barabasi_emergence_1999}, its elegant model does not translate into real life applications.

Albert and Barabasi \cite{barabasi_emergence_1999} theorized the scale-free graph to address the discrepancies between real systems and the properties of random graphs. Scale-free graph is generated by having $N$ connected nodes. When node $(N+1)$ gets introduced, it gets connected to each other node with a probability proportional to node’s properties, as shown in algorithm \hyperlink{Algorithm_1}{2} at table \ref{tabtable_1}.

\begin{table}[tb]
\centering
\caption{Link creation algorithms}
\begin{tabular}{|c|c|c|}
\hline
\begin{minipage}{1.36in}
\hypertarget{Algorithm_1}{}
 \lstset{language=Pascal} 
 \begin{lstlisting}[
    basicstyle=\footnotesize, %or \small or \footnotesize etc.
]
Algorithm 1:
Add_node_random(N+1)
{
 Foreach (node i in N)
   {
     r = Random(0, 1)
     if( r < p)
       link(i, N+1)=1
   }
}
\end{lstlisting}

\end{minipage}

&

\begin{minipage}{1.5in}
\label{algo2}
 \lstset{language=Pascal} 
 \begin{lstlisting}[
    basicstyle=\footnotesize, %or \small or \footnotesize etc.
]
Algorithm 2:
Add_node_scale_free(N+1)
{
 Foreach (node i in N)
   {
     r = Random(0, 1)
     if( r < pi)
        link(i, N+1)=1
   }
}
\end{lstlisting}
\end{minipage}
&

\begin{minipage}{1.5in}
 \lstset{language=Pascal} 
 \begin{lstlisting}[
    basicstyle=\footnotesize, %or \small or \footnotesize etc.
]
Algorithm 3:
Add_node_evol_rand(N+1)
{
 Foreach (node i in N)
  {
    r = Random(0, 1)
    if( r < p/N)
       link(i, N+1)=1
  }
}

\end{lstlisting}
\end{minipage}
\\
\hline
\end{tabular}\label{tabtable_1}
\end{table}

The BitOverNet is free from spatial-temporal restrictions imposed on physical networks and from human-introduced biases such as popularity in the case of WWW. It can be modeled as having $N$ connected nodes, when node $(N+1)$ gets introduced, it gets connected to each other node with the same probability $p_{(N+1)}=\frac{p}{N}$, as shown in algorithm \hyperlink{Algorithm_1}{3} at table \ref{tabtable_1}. It is similar to random networks in the way that a new node connects equiprobably to all existing nodes, but as more nodes join the network, the probability of connection decreases. In bitcoin, each new node has to establish a fixed number of outgoing connections called $m$. It is worth noting that for implementation purposes, every released "Bitcoin Core" version has a hard-coded list of IP addresses that were available during the time the specific version was released \cite{Bitcoinwebsite}. This list does not create any guarantees about the availability of those addresses, else this centrality results in a fatal flaw.

\begin{figure}
    \centering
    \includegraphics[scale=0.5]{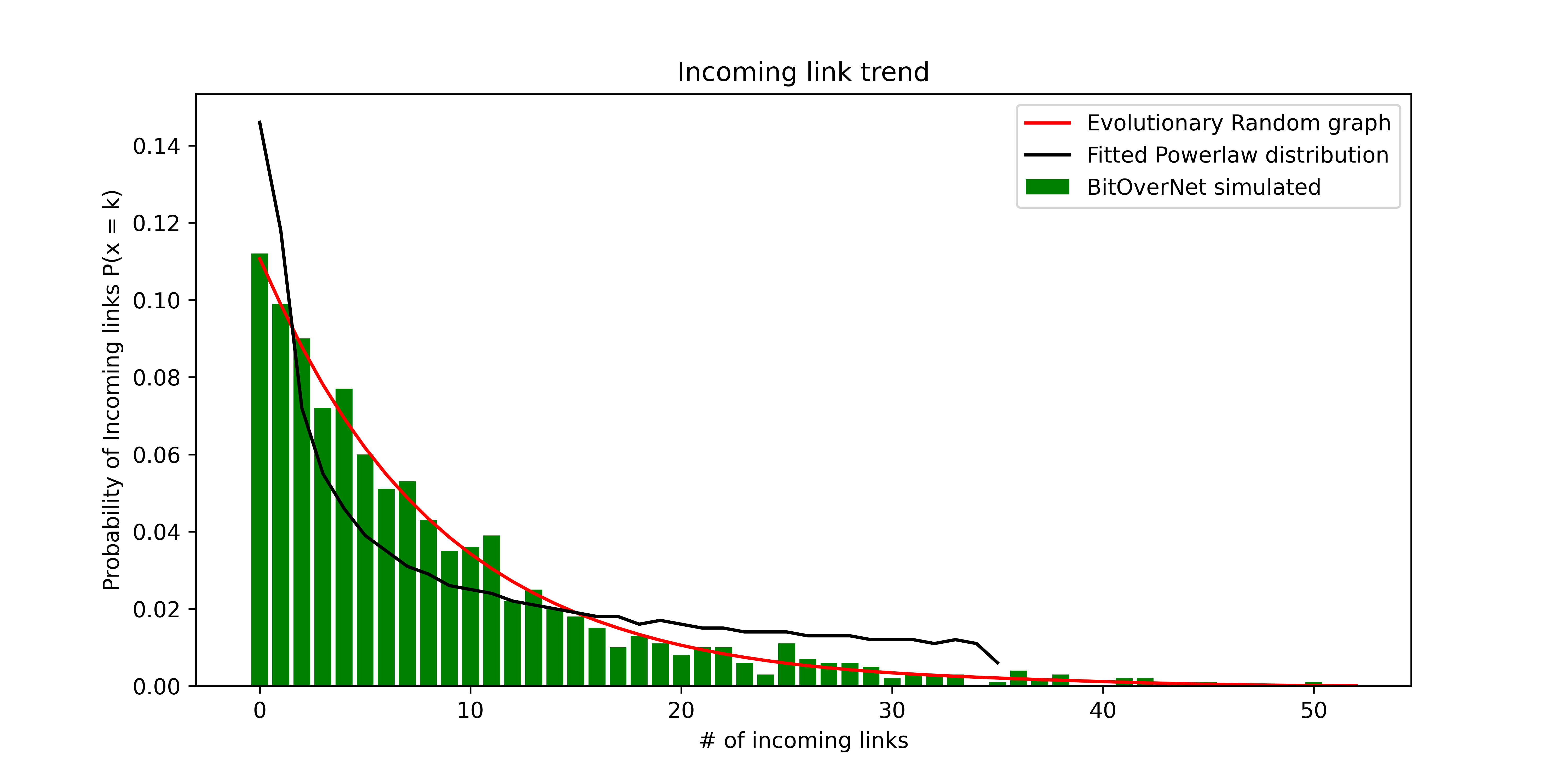}
    \caption{BitOverNet's incoming link distribution}
    \label{fig:evolutionary-random}
\end{figure}

We propose "Evolutionary Random" graph as a theoretical model that describes how BitOverNet behaves, as shown in algorithm 3, having a probability density function (expected portion of nodes having $k$ incoming links): 
\begin{equation}
    P(x=k) = \sum_{i=1}^{N} \frac{m^k (H_N - H_{max(m,i)})^k}{k! \times e^{m(H_N - H_{max(m,i)})}} 
\end{equation}
Where $H$ is the harmonic number. To measure the accuracy of our bitcoin model, we simulated a BitOverNet (of 1000 miners) and measured the number of peers having $k$ links. Figure \ref{fig:evolutionary-random} shows the results of the bitcoin network in addition to the predicted bitcoin model and a fitted power-law distribution. It can be easily shown that BitOverNet does not show scale-free properties and that evolutionary-random graph nicely predicts the dynamics of the bitcoin network. We can now utilize evolutionary-random graph as a foundation for understanding the BitOverNet, predicting forks and considering its implications.

\section{Network Diameter}\label{sec2}
Network diameter has been defined by \cite{albert_statistical_2002} as “the maximal distance between any pair of its nodes.” Random graphs have, in average, the diameter \cite{albert_statistical_2002, chung_diameter_2001}:
\[ d_{random} = \frac{ln(N)}{ln(pN)} =  \frac{ln(N)}{ln(\langle k \rangle)}
\]
where $k$ is the number of links per node and $\langle k \rangle$ is the average number of links. Scale-free networks scale better than random graphs when it comes to diameter which can be represented as \cite{albert_statistical_2002}:
\[ d_{scale-free} = A \times ln(N - B ) + C
\]
where $A$, $B$ and $C$ are fitting configuration parameters. Its diameter tends asymptotically to
\[ d_{scale-free} \propto \begin{cases} \frac{ln(N)}{ln(ln(N))} & \lambda = 3, \mbox{   (Bollobas and Riordan \cite{bollobas_diameter_2004})} \\ ln(N) & \lambda > 3,\mbox{     (Cohen and Havlin \cite{cohen_scale-free_2003})} \end{cases}
\]

BitOverNet's diameter can be calculated theoretically, following evolutionary random graph, as:

\begin{equation}
d_{evolutionary-random} = \frac{ln(N -2 \times m )}{ln(\sum_{k=1}^{\infty} \frac{(k + m) \times m^k}{k!} (\sum_{i=1}^{N} \frac{(H_N - H_{max(m,i)})^k}{e^{m(H_N - H_{max(m,i)})}})  )} +2
\end{equation}

\begin{figure}
    \centering
    \includegraphics[scale=0.5]{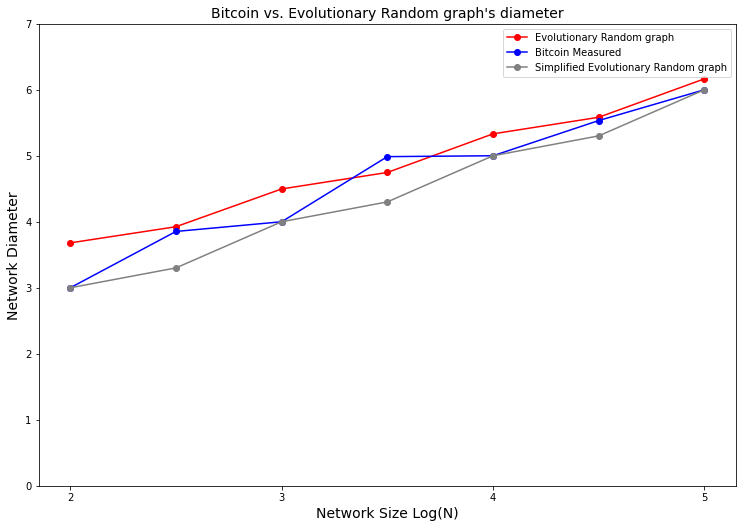}
    \caption{BitOverNet vs. Evolutionary Random graph's diameter}
    \label{fig4.a}
\end{figure}

To test the predicted network’s accuracy, we simulated the bitcoin network with different sizes (values of $N$ between $100$ and $100000$) and measured its network diameter averaged over 100 iterations. The values of the measured and predicted bitcoins can be seen in figure \ref{fig4.a}. It is observable that the diameter scales with factor of $10$ (linear over log axis), so we can present a simplified version of the evolutionary random graph's diameter before elaborating on its scalability. The evolutionary random graph's diameter can be simplified as follows:

\begin{equation}
    simplified-d_{evolutionary-random} = 1 + log(N) 
\end{equation}

\begin{figure}
    \centering
    \includegraphics[scale=0.5]{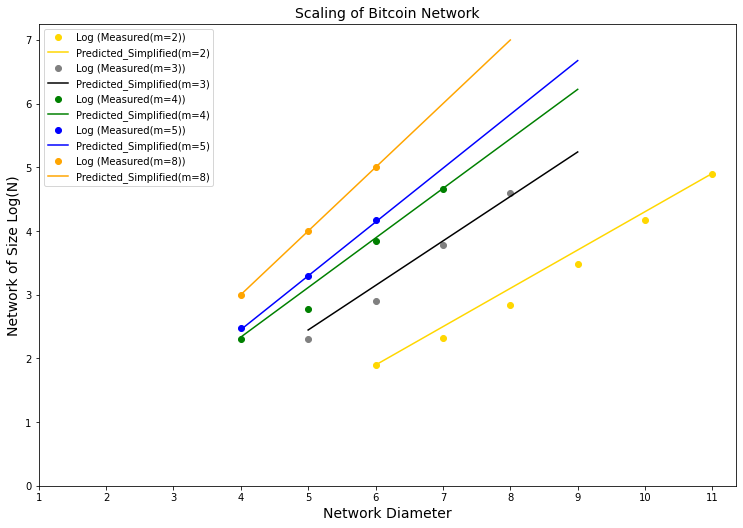}
    \caption{Scaling of Evolutionary Random graph}
    \label{fig4.b}
\end{figure}

To verify whether scaling by factor of $10$ was serendipitous, we simulated (following the above configuration) the bitcoin network for different values of $m$ (number of outgoing connections). We found that it scales by factor of “$m+2$”. Since bitcoin’s default $m$ is $8$, it scaled by factor of 10. Accordingly, we can generalize the simplified evolutionary random graph's diameter, as shown in figure \ref{fig4.b}, into:

\begin{equation}
  simplified-d_{evolutionary-random} = $1$ + log_{m+2}(N) + \epsilon
\end{equation}

The correctness of BitOverNet's diameter model has been validated, using measured data, in section \ref{result_verification_block_propagation}.

\subsection{Result verification: Block propagation delay}\label{result_verification_block_propagation}
\begin{figure}
    \centering
    \includegraphics[scale=0.5]{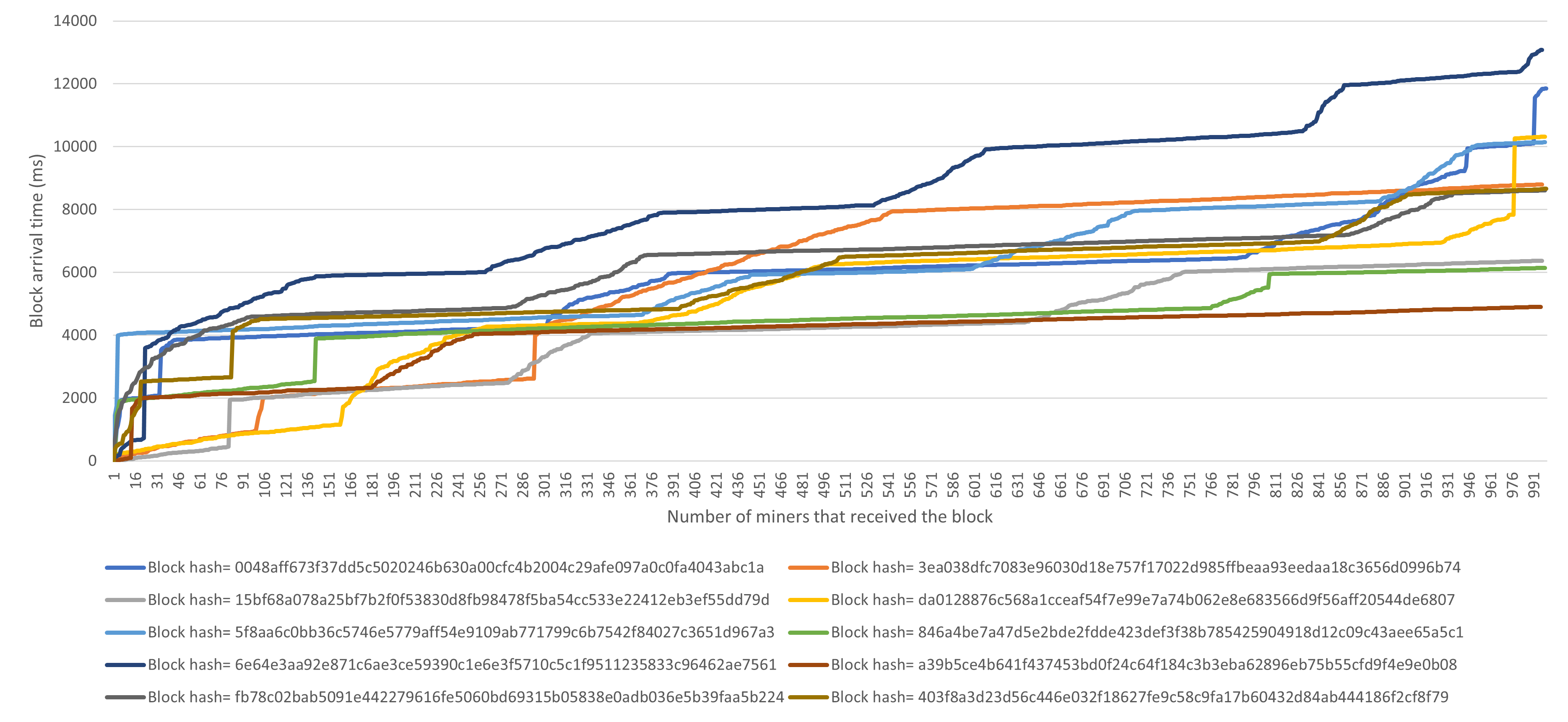}
    \caption{Block propagation (first 1000 nodes)}
    \label{fig5.a}
\end{figure}

\begin{figure}
    \centering
    \includegraphics[scale=0.5]{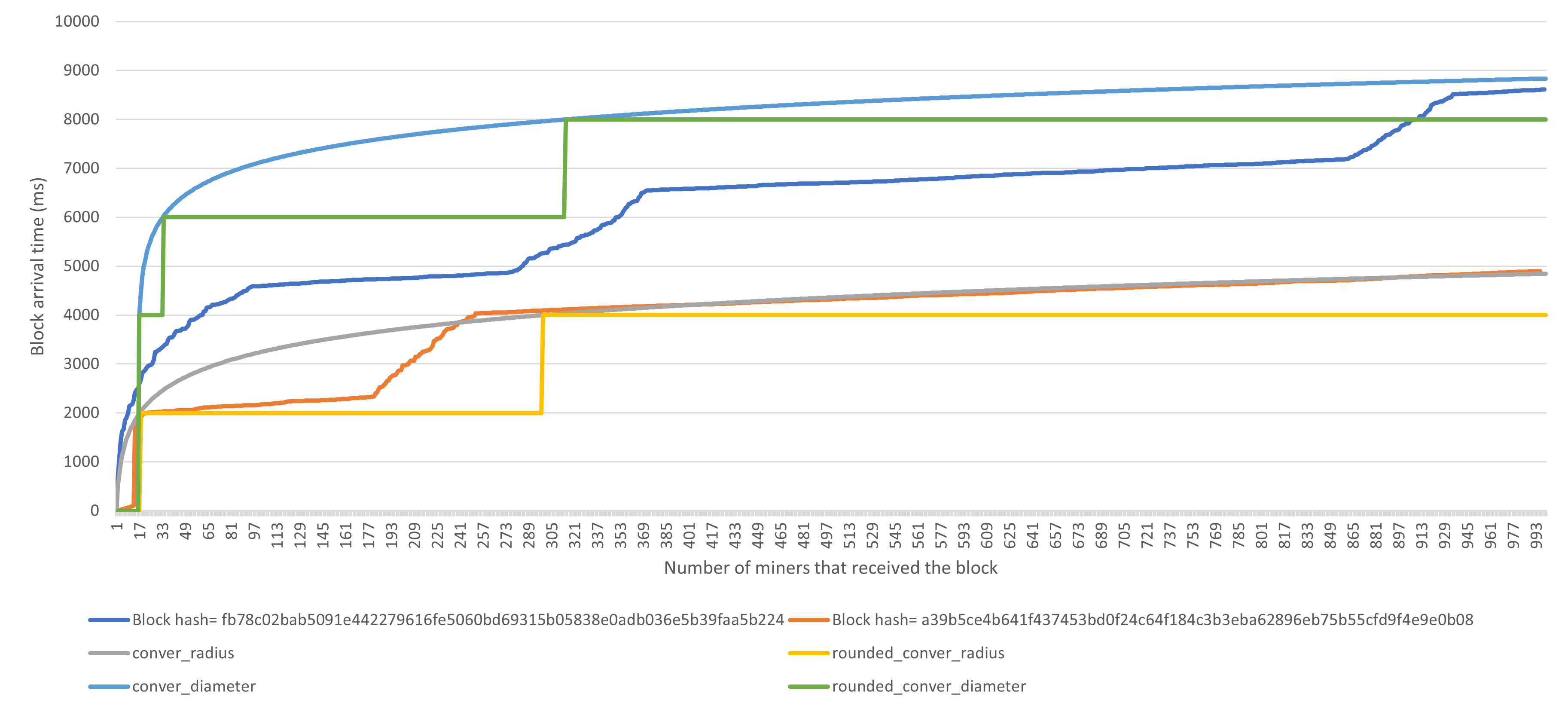}
    \caption{Block propagation vs. prediction}
    \label{fig5.b}
\end{figure}

Blocks generated by miners located at the periphery of the network (nodes with low centrality) need $d$ hops to reach all the networks, where $d$ is the network diameter. The blocks generated by miners located at the center of the network (miners with high centrality) need less hops than those starting from the periphery (let us call it network radius). Blocks generated by other miners need a number of hops between network radius and diameter. It is expected that the convergence of those blocks starting from the center follows a pseudo logarithmic shape, while those starting from the periphery follows a pseudo sigmoidal shape.
Following $d_{evolutionary-random}$, we can deduce the convergence (propagation) of a block based on its source. Accordingly, we define the convergence of blocks starting at the center and at the periphery as follows:

\begin{equation}
    conver_{radius} = \frac{ln(n)}{ln(\sum_{k=1}^{\infty} \frac{(k + m) \times m^k}{k!} (\sum_{i=1}^{N} \frac{(H_N - H_{max(m,i)})^k}{e^{m(H_N - H_{max(m,i)})}})  )} \times Shd 
\end{equation}

\begin{equation}
    conver_{diameter} = (\frac{ln(n -2 \times m )}{ln(\sum_{k=1}^{\infty} \frac{(k + m) \times m^k}{k!} (\sum_{i=1}^{N} \frac{(H_N - H_{max(m,i)})^k}{e^{m(H_N - H_{max(m,i)})}})  )} +2) \times Shd
\end{equation}

where n is the number of miners the block reached. Full convergence is achieved when $n = N$ . Shd is the "Single hop delay" which is equivalent to the block's link propagation between two individual miners and the time needed for the receiving miner to verify its correctness. The first $1000$ miners receiving a block are recorded in \cite{dataset1, dataset2} and this provides a great opportunity to verify our predictions. The correctness of the predictions have been tested for all blocks, but we are only showing illustrative samples. It can be seen in figure \ref{fig5.a} that the Shd is around 2000ms since the block arrival time is varying in multiples of 2000ms. Some of the blocks (such as the block with hash starting with a39b5ce4b64...) reach a considerable number of miners in the first Shd and this represents blocks generated by miners with high centrality. Other blocks (such as block with hash starting with  fb78c02...) reach a small number of miners in the first Shd and this represents blocks generated by miners with low centrality. It can be seen in figure \ref{fig5.b} that the convergence equations describe the block propagation of blocks generated by a node with high/low centrality.

\begin{figure}
    \centering
    \includegraphics[scale=0.4]{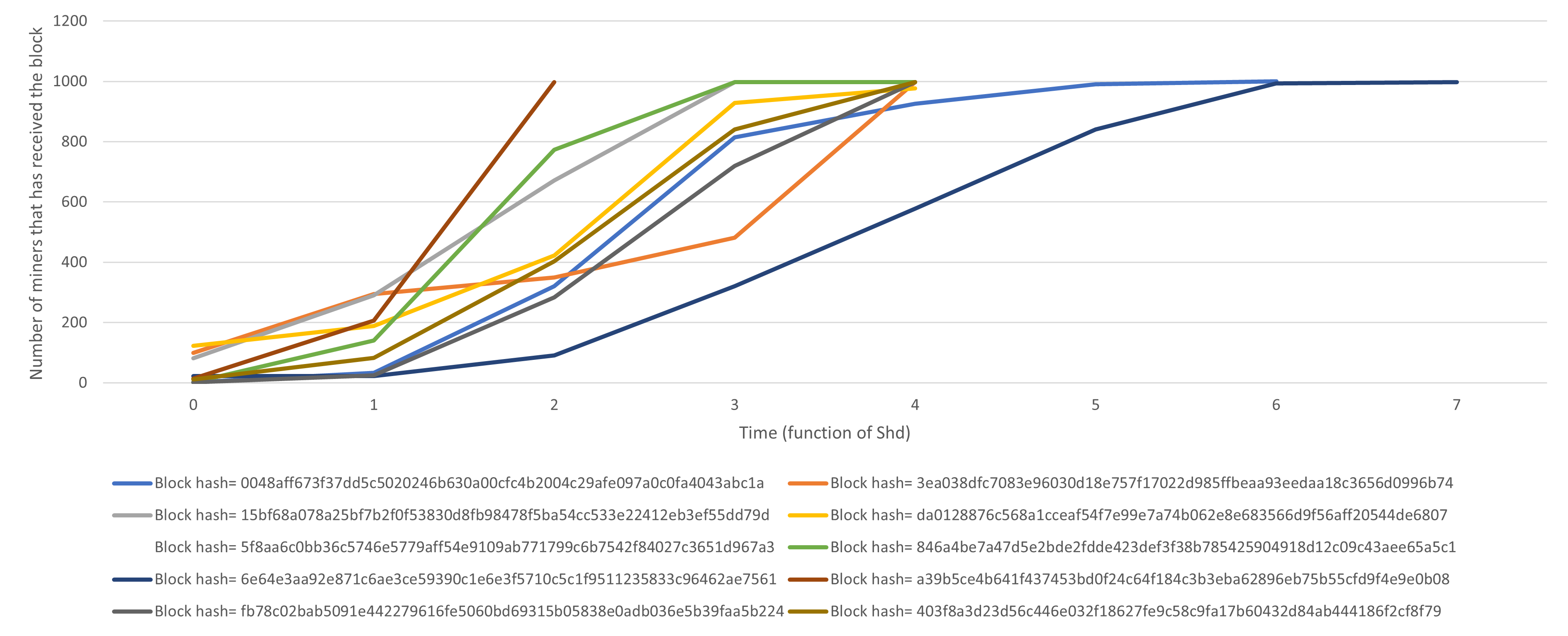}
    \caption{Cumulative block propagation (first 1000 nodes)}
    \label{fig:Inv_block_arrival}
\end{figure}

Using the same data, but showing the number of blocks reached at every Shd, as shown in figure \ref{fig:Inv_block_arrival}, it can be seen that the block issued by a node with high centrality shows a pseudo logarithmic convergence (only the first part is shown since only the arrival time at the first 1000 nodes is logged). Similarly, it can be seen that the block issued by a node with low centrality shows a pseudo logarithmic convergence.


\section{Forking Probability and Mining Difficulty}\label{sec13}
As discussed at the start of this work, forking happens when the inter-arrival time between two mined blocks is less than the propagation delay. It is not a simple Poisson distribution because when a miner receives a block, from its neighbor, it stops mining and waits for consensus. This converts the problem into a sequence of Poisson distributions each with less rate $\lambda$ (since after every period, less miners remain mining, uninformed of the mined block, until reaching $0$ and this when consensus is achieved). The forking probability is then:

\begin{equation}
\begin{split}
1- \prod_{i=1}^{\lceil d_{evol-rand} \rceil} e^{-shd \times \lambda_{mine}\times \psi^i} \le P(forking) \le
1- \prod_{i=1}^{\lfloor d_{evol-rand} \rfloor} e^{-shd \times \lambda_{mine}\times \psi^i}
\end{split}
\end{equation}\\
where  $\psi = (\sum_{k=1}^{\infty} \frac{(k + m) \times m^k}{k!} (\sum_{i=1}^{N} \frac{(H_N - H_{max(m,i)})^k}{e^{m(H_N - H_{max(m,i)})}}))$, shd is the propagation
delay between two miners in addition to the time a miner takes to verify the correctness of the generated block and $\lambda_{mine}$ is the probability of one miner generating a block in a period of time. $\lambda_{mine}$ can be simplified as:
\begin{equation}
    \lambda_{mine} = \frac{\mbox{computational-speed}}{\mbox{mining-difficulty}}
\end{equation}
Knowing the forking probability, we can change the mining difficulty proactively in a way that forking probability remains below set $threshold$. This is an important upgrade to the reactive way bitcoin has been and is currently being managed. Radically increasing the mining difficulty can drop the forking probability but causes mining to become financially infeasible and this depletes bitcoin of its miners. The developed model helps in calibrating the network parameters. The minimum mining difficulty for keeping forking probability below $threshold$ is:

\begin{equation}  
\mbox{mining-diff}  \ge \frac{-\mbox{computational-speed} \times \sum_{j=1}^{\left\lfloor d_{evol-rand} \right\rfloor} shd \times \Psi^{j}}{ln(1 - threshold)}
\end{equation}

\section{Economic Equilibrium}\label{sec13}
The BitOverNet will remain stable if the miners are making enough profits to keep them running, but not to attract additional miners which will decrease the profits generated by all other miners. This is a two-stage equilibrium problem, the first is within the same network diameter where additional miners cause the mining profits to be divided over a larger pool of miners and thus resulting in less profits per miner. The second stage is when the increase in the number of miners results in drastic increase in the network diameter. The first stage is a classic economics problem, so we will only focus on the second. Let us consider the profit of mining a block to be $\mbox{Profit-mining}$. The profit of mining a block is function of the bitcoin price which is not our interest in this paper and will only be represented as $\mbox{profit-mining}$. Let the cost of mining be $\mbox{cost-mining}$:
\begin{equation}
    \mbox{cost-mining} = \frac{c \times \mbox{mining-difficulty}}{\mbox{computational-speed}}
\end{equation}
\\
where $c$ is a constant related to costs (such as electricity). Let $\mbox{computational-speed}$ be the average computational speed per miner. For miners to remain interested in mining, $\mbox{profit-mining}$ – $\mbox{cost-mining}$ should remain positive and thus the number of miners should remain:
\begin{equation}
    N \le \frac{\mbox{profit-mining}}{\mbox{c}} \times \frac{-ln(1 - threshold)}{\sum_{j=1}^{\lfloor d_{evol-rand} \rfloor} shd \times \psi^i}
\end{equation}

Using the simplified network diameter, we can conclude that:
\begin{equation}
    N \times (1 + log(N)) \times e^{\frac{ln(N -2 \times m)}{log(N) -1}} \le \frac{-ln(1 - threshold) \times \mbox{profit-mining}}{c}
\end{equation}

Equilibrium is achieved when profit =0 and thus the stable network size is:
\begin{equation}
    N \times (1 + log(N)) \times e^{\frac{ln(N -2 \times m)}{log(N) -1}} = \frac{-ln(1 - threshold) \times \mbox{profit-mining}}{c}
\end{equation}
\section{Conclusion}\label{sec13}

The contributions of this work are multi-folded. First, we developed a theoretical model for bitcoin overlay network to overcome the technical hurdles facing mapping it. Second, we calculated the minimum mining difficulty required for limiting forking. As can be seen in \cite{neudecker_network_2019}, the forking rate dropped significantly after July $2018$, and this is not healthy since it was due to a radical increase in mining difficulty. 

Bitcoin mining is an investment which makes it dependent, not just on bitcoin's price, but on involved risks, revenues from alternative investments, legal constraints, economic growth and tolerance to risk. In a future work, we will investigate profit-mining and incorporate it as investment attractiveness \cite{elton2009modern}.

\bibliography{sample}

\section*{Author contributions statement}

J.B.A. designed the mathematical model, J.B.A. and S.D. developed the simulation, J.B.A. and B.Q. collected the measurements, J.B.A. and L.H. analysed the results. All authors reviewed the manuscript. 

\section*{Additional information}

\textbf{Competing interests}\\ 
The author(s) declare no competing interests.

\end{document}